\documentclass[lettersize,journal]{IEEEtran}
\usepackage{cite}
\usepackage{amsmath,amssymb,amsfonts}
\usepackage{algorithmic}
\usepackage{algorithm}
\usepackage{graphicx,color}
\usepackage{soul}
\usepackage{hyperref} 
\usepackage{cleveref}

\usepackage{textcomp}
\usepackage{booktabs}
\usepackage{array}

\usepackage{xcolor}

\begin{document}

\renewcommand{\thesubsection}{\thesection\Alph{subsection}} 
\renewcommand{\thesubsubsection}{\thesubsection\arabic{subsubsection}} 
\numberwithin{subsection}{section} 
\numberwithin{subsubsection}{subsection} 

\newcommand{\pavan}[1]{{[\textcolor{blue}{#1}]}}
\newcommand{\verify}[1]{{[\textcolor{red}{#1}]}}
\newcommand{\mcnair}[1]{{[\textcolor{red}{McNair comment: #1}]}}

\title{Deep reinforcement learning-based optimization of multi-RAT handovers}
\author{
Dr. Pavan Kumar Mangipudi, Satya Krishna Pothapragada, Dr. Alina Zare, Dr. Janise McNair \\

\vspace{2mm}
Department of Electrical and Computer Engineering \\
University of Florida, Gainesville  USA\\
   \tt
   \small mspkumar.11@gmail.com, \{spothapragada,azare,mcnair\}@ufl.edu

\thanks{Manuscript received XX Month, XXXX; revised XX Month, XXXX; accepted XX Month, XXXX. Date of publication XX Month, XXXX; date of current version XX Month, XXXX.}%
\thanks{This material is based upon work supported by the National Science Foundation under Grant Numbers 2030122 and 1738420.}}

\markboth{Deep Reinforcement Learning-based Handovers in Multi-RAT networks}{Author \textit{et al.}}

\maketitle

\begin{abstract}
A major evolutionary goal and challenge for next-generation wireless systems is seamless multiple radio access technology (multi-RAT) mobility. In a multi-RAT environment, handover decisions must account for both user quality of service and network priorities, including signal quality, resource availability, energy consumption, latency, and security. Although WiFi/cellular mobility is well explored in research and standardization, many conventional handover mechanisms continue to rely primarily on received signal strength. Such decisions may overlook link delay, congestion, and the distinct operating characteristics of heterogeneous RATs. In this work, we present the Reinforced Flexible Multi-RAT Selection framework (ReFlexMRS), a configurable Deep Reinforcement Learning (DRL)-based approach for multi-criteria handover target selection. ReFlexMRS is evaluated using an SDN-enabled multi-RAT testbed. The decision problem is formulated as a Markov Decision Process with a finite action space and is addressed using a Deep Q-Network (DQN) with convolutional layers that extract relationships among network measurement features. The agent is trained offline from collected network observations using reward guided Q-value updates, then used to select handover targets during inference. A 5G-WiFi case study demonstrates that ReFlexMRS can trade a small amount of throughput for substantial improvements in link delay, handover stability, and WiFi offloading. Compared with RSSI-based selection, the proposed formulations retain approximately 90--95\% of the baseline throughput while reducing total handovers by up to 18\% and WiFi-to-5G handovers by up to 13\%, while maintaining the required RSSI levels.
\end{abstract}


\section{INTRODUCTION}\label{sec_intro}

\IEEEPARstart{T}{hroughout} the evolution of wireless and cellular networks, improving coverage and connectivity has remained a central objective, driven by the rapid growth of mobile users, connected devices, and data intensive applications. This trend is expected to intensify in upcoming 6G cellular networks, where systems must support higher coverage and connectivity demands, denser user populations, and the reliable integration of Internet of Things (IoT), satellite, high altitude platform, WiFi, and other heterogeneous users~\cite{wang20236gsurvey}. A key requirement for 6G will therefore be the development of technologies and protocols that enable coordination, reduce operational complexity, and support mobility within and across overlapping heterogeneous multiple radio access technology (multi-RAT) environments~\cite{mcnair20226g,pan2025biglitemultirat}. Research on multi-RAT networking aims to enable users to move seamlessly among diverse access networks while leveraging heterogeneous services to maintain reliable quality of service (QoS)~\cite{mangipudi2023survey,ullah2025aimobility}.

\begin{figure}[ht!]
\centerline{\includegraphics[width=\columnwidth]{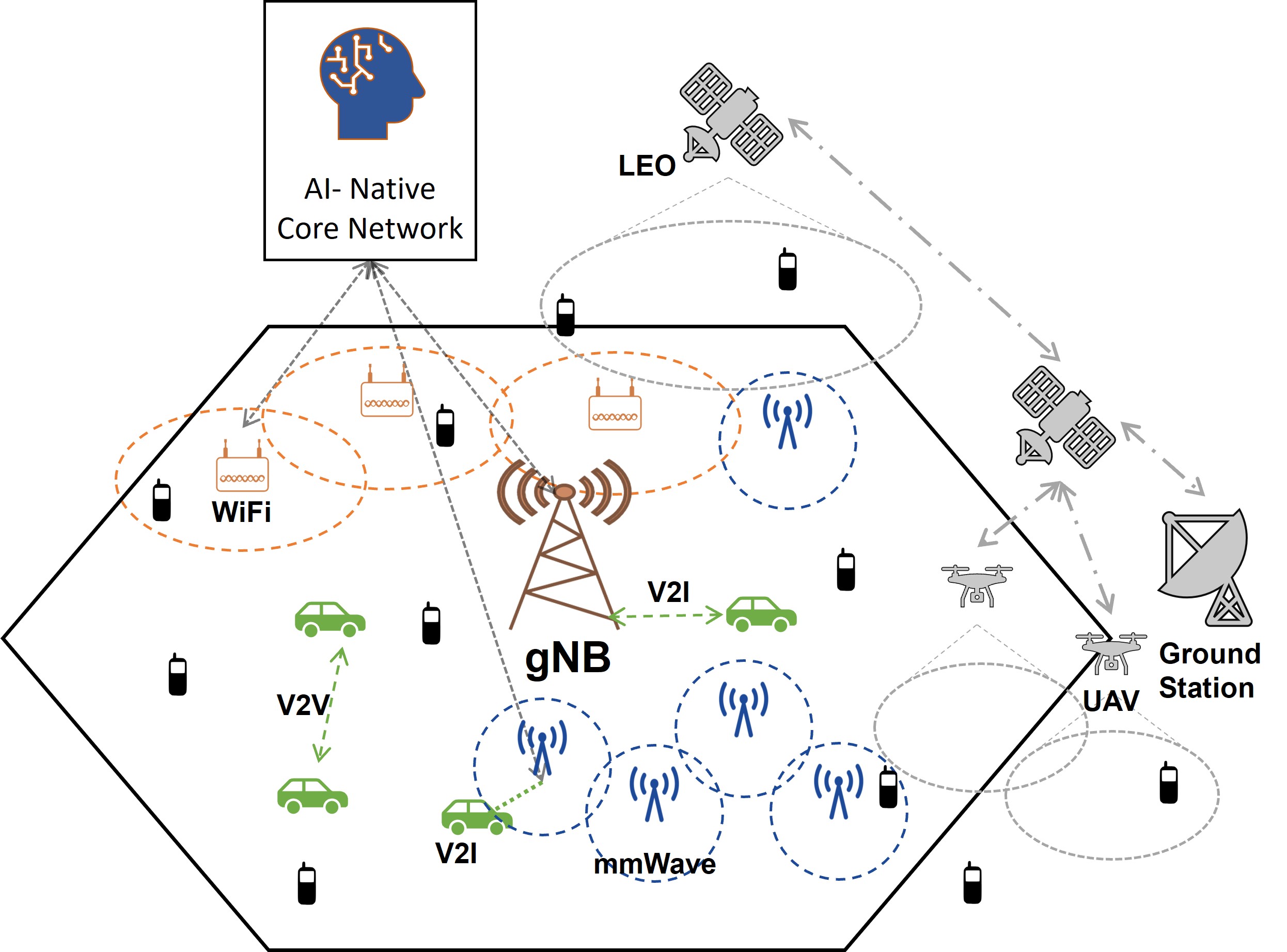}}
\caption{A high-level 6G multi-RAT network topology.}
\label{fig:6G_multiRAT}
\end{figure}

Although the integration of WiFi and cellular networks has been widely explored in both research and standardization, several emerging multi-RAT scenarios remain unresolved. Existing standards have addressed specific forms of interworking, such as LTE-WiFi aggregation in 4G, Access Traffic Steering, Switching, and Splitting (ATSSS), and the non-3GPP Interworking Function (N3IWF) in 5G~\cite{etsi5Gwithsdn,pieska2024atsss}. However, recent advances in wireless standards and the growing demand for diverse access technologies have created an increasingly overlapping wireless ecosystem~\cite{charfi2025wifi7to8}, illustrated in Fig.~\ref{fig:6G_multiRAT}. For example, WiFi 6E~\cite{standard:WiFiax} extends WiFi operation into the 6 GHz band, creating new opportunities and challenges for coexistence with cellular deployments. Similarly, 5G New Radio Unlicensed (NR-U) allows 5G systems to operate in unlicensed bands traditionally used by WiFi, including the 5 GHz and 6 GHz bands~\cite{Hirzallah2021nru}.

Other technologies, including vehicular networks using vehicle-to-vehicle (V2V) and vehicle-to-infrastructure (V2I) protocols such as IEEE 802.11bd~\cite{standard:WiFiV2X}, networked satellites, unmanned aerial vehicles (UAVs), and IoT systems, may also coexist with or be integrated into 5G and WiFi environments~\cite{mahboob2024satellitentn}. Furthermore, other wireless services, such as radio astronomy, space research, and mission-oriented Low Earth Orbit (LEO) satellite systems, may compete for shared spectrum and network resources~\cite{mcnair20226g}. This ecosystem introduces significant challenges related to interference, coordination, coexistence, and resource allocation. Conversely, it also creates opportunities to reduce congestion and latency, improve QoS, and adapt security policies by dynamically assigning users to the most suitable access networks. Realizing these benefits requires advanced coordinated multi-network management and adaptive multi-RAT handover mechanisms. However, limited research and few standard implementations currently address seamless handover management across general coexisting multi-RAT networks.

Software-defined networking (SDN) has emerged as an effective architecture for network management and control, and it provides a promising framework for multi-RAT network coordination~\cite{mangipudi2023survey}. By separating the control and data planes, SDN enables dynamic network reconfiguration, centralized policy enforcement, and flexible resource optimization. These capabilities can help address operational complexity, limited interoperability, scalability, and adaptive resource utilization in heterogeneous wireless environments. SDN can also be combined with artificial intelligence and machine learning (AI/ML) techniques within beyond-5G (B5G) and 6G architectures. For example, the O-RAN Alliance introduces RAN Intelligent Controllers (RICs), building on 3GPP NR specifications, to support programmable and intelligent RAN control~\cite{polese2023understanding}. Through RICs, AI/ML-based techniques can optimize handover decisions using measurement report parameters such as offset, hysteresis, time-to-trigger, and other newly defined criteria~\cite{3gpp2020nrrrc}.

6G networks are expected to be AI-native, implying comprehensive integration of AI into the protocol stack. This approach contrasts with using AI for specific tasks alone, as an AI-native protocol stack would manage all cellular and wireless system tasks through AI-based implementations and corresponding new sets of protocols. This paradigm shift towards AI necessitates that the development of multi-RAT handover algorithms can and should employ AI-based techniques that can generalize across diverse deployments and application scenarios. We note that supervised learning algorithms may be less suitable for this task as they depend heavily on training data from specific network conditions. Reinforcement Learning (RL), in contrast, offers a promising approach because it learns through interaction with the environment and is less dependent on manually labeled data. In RL, the problem is commonly formulated as a Markov decision process (MDP), consisting of states, actions, and rewards. This formulation is well suited to handover optimization, as demonstrated by prior studies on RL-based handover in individual cellular, satellite, and WLAN environments~\cite{wang2021DRLsatellite,han2019a,tan2022a}. RL has also been applied to multi-RAT access, network selection, resource allocation, and traffic steering~\cite{yacheur2023drlrat,allahham2022multiagent,desantis2022satellite,Habib2023trafficsteering,habib2023hierarchical}.  However, comparatively fewer works provide a methodology for adaptive mobility-driven multi-RAT handover target selection while examining how a common learning architecture can accommodate both general QoS objectives and RAT-specific operator preferences. This distinction matters because a target selection policy intended for one heterogeneous deployment may need different reward constraints or priorities in another, even when the overall decision framework is unchanged.

In this work, we propose ReFlexMRS, a configurable deep reinforcement learning (DRL)-based framework for multi-criteria handover target selection across heterogeneous access networks, and evaluate it in an SDN-enabled multi-RAT environment. The target selection decision is formulated as an MDP with a finite action space and learned using a Deep Q-Network (DQN) architecture. Additionally, the model employs convolutional layers to process correlations among measurements collected from users and candidate access nodes. Despite the growing body of literature on AI-driven handover optimization in multi-RAT networks, several gaps remain. First, much of the handover literature still relies heavily on signal strength metrics such as received signal strength indicator (RSSI) or reference signal received power (RSRP). Although stronger signal levels often correlate with improved throughput, signal strength alone does not capture other modern factors such as security level, energy consumption, in addition to latency and other distinct operating characteristics of overlapping heterogeneous RATs. Second, many AI-based approaches address either single-RAT handover or heterogeneous access association, resource allocation, and traffic steering, but not mobility-driven target selection across cellular and local access networks. Third, limited consideration has been given to comparing general, network independent reward formulations with deployment specific formulations that incorporate RAT characteristics and operator-defined preferences.

To address these gaps, we present ReFlexMRS, a reinforcement learning approach that co-optimizes handover decisions across heterogeneous access networks using multiple performance criteria beyond RSSI. Specifically, our contributions are as follows:
\begin{itemize}
    \item We develop a DRL-based handover mechanism that can process varying sets of decision criteria, providing greater flexibility in target network selection than conventional RSSI-based handover approaches. Offline reward guided DQN training learns a handover target selection policy from collected network observations, rather than relying solely on a manually specified, fixed weighted score

    \item We propose ReFlexMRS, a unified DRL-driven framework for mobility across heterogeneous access networks. In the considered  case study, a user equipment (UE) can choose among a set of multiple local access points and a gNB node while the RL agent learns the handover target selection policy under varying network conditions and mobility patterns.

    \item We develop both deployment specific and network independent ReFlexMRS formulations to demonstrate the flexibility and adaptability of DRL-based handover. By incorporating RAT-specific information, the deployment specific formulation can prioritize selected categories of access nodes and support local access network offloading when the configured signal quality and delay conditions are satisfied.

    \item We evaluate ReFlexMRS in a case study of heterogeneous 5G-WiFi network nodes using link delay, throughput, total handovers, and WiFi-to-5G handovers, thereby characterizing both QoS and handover stability trade-offs relative to conventional RSSI-based selection.
    \item We formulate target selection for an arbitrary number of candidate access nodes and measurement criteria, with configurable combined or separate state features and reward rules. The formulation can therefore be instantiated for other heterogeneous RAT combinations, although performance is evaluated here only for the specified 5G-WiFi case study.
\end{itemize}

This study evaluates the handover decision layer in a single controlled 5G-WiFi emulation, with offline training and an RSSI-based baseline. The results provide a proof of concept for the framework and illustrate QoS and handover count trade-offs in this setting. Full protocol stack operation, performance in other deployments, and cross-RAT model transfer remain outside the experimental scope and are considered future validation tasks.

In section \ref{sec_related_work}, we present an overview of related work in handovers in cellular and multi-RAT networks. We also examine recent literature on handover decision criteria optimization and RL-based handovers in wireless and cellular networks. Section \ref{sec_background} provides an overview of the enabling technologies discussed in this work, including SDN, handovers, and RL algorithms. Section \ref{sec_methodology} defines the general ReFlexMRS target selection framework, and Section \ref{subsec_env_desc} describes its implementation in the 5G-WiFi case study. Section \ref{sec_results} details the experimental results. Sections \ref{sec:discussion} and \ref{sec_conclusion} provide a discussion followed by the conclusion of the paper.

\section{Related Work}\label{sec_related_work}
When a wireless or mobile connection becomes untenable due to channel or received signal degradation, handover is the process that transfers the wireless connection from one channel or cell to another, with the goal that there is no drop in the connection experience during this movement. While handovers in wireless and cellular networks have been extensively researched, this paper focuses on the comparatively less explored problem of multiple radio access technology (multi-RAT) handover.

The vast majority of conventional multi-RAT handover mechanisms use received signal strength indicators (RSSI) as a principal criterion for triggering or selecting a handover. However, significant differences in RAT characteristics enable target network selection based on additional system or user constraints. For example, a millimeter-wave (mmWave) 5G network can provide high data rates and low latency, but its higher frequency signals are more susceptible to blockage and attenuation. WiFi networks, while having different range and capacity characteristics, can provide strong indoor coverage. Thus, selection based only on signal strength may overlook QoS criteria such as link delay that affect a user's experience. Furthermore, users may have a variety of options to access the network, where the availability of each option may change over time. Predefined decision criteria may be insufficient when the available access nodes and QoS conditions change with mobility

Multi-criteria decision-making (MCDM) based handover schemes have been proposed in recent years~\cite{ciciouglu2021multi,rizkallah2018sdnverticalho,mangipudi2024mcdm}. These approaches incorporate additional criteria that influence user QoS while maintaining minimum signal strength requirements. However, MCDM methods typically require predefined weights or thresholds that must be customized for a particular network configuration and propagation environment. AI-based handover methods have also been proposed for individual RAT environments, including WLANs~\cite{han2019a} and cellular networks~\cite{haldorai2019a,tan2022a,mollel2020doubledrl,mollel2021survey}. These studies demonstrate that learning-based policies can improve handover decisions under dynamic mobility and channel conditions, but their decision spaces generally contain access nodes belonging to the same type of RAT.

As an alternative to MCDM or other optimization, several works have proposed Deep Reinforcement Learning (DRL)-based handover algorithms to address QoS and reduce frequent handovers, particularly in dynamic vehicular and cellular environments. RL can dynamically learn and adapt to the network environment by iteratively assessing the quality of handover decisions and refining its policy. Tan \textit{et al.}~\cite{tan2022a} apply Double Deep Q-Learning to handover decisions based on RSRP measurements from vehicles moving at high speed. Their approach remains primarily driven by signal strength measurements within a cellular network. Similarly, Mollel \textit{et al.}~\cite{mollel2020doubledrl} employ Double Deep Q-Learning to reduce frequent handovers and improve QoS in a mmWave cellular network, with SINR as the primary decision metric. Yajnanarayana \textit{et al.}~\cite{yajnanarayana2020a} model 5G handover as a contextual multi-armed bandit problem and use Q-learning to optimize mmWave beam selection based on measurements related to signal strength. Han \textit{et al.}~\cite{han2019a} use a convolutional and recurrent DQN for handoff management in dense WLANs, extracting temporal features from SINR measurements and optimizing uplink throughput. Collectively, these studies demonstrate the effectiveness of RL for handover optimization, but they do not address mobility-driven target selection across heterogeneous RATs while jointly considering RSSI, delay, handover stability, and RAT-aware offloading.

Several studies have extended RL from single-RAT handover optimization to heterogeneous network selection and multi-RAT traffic management. Yacheur \textit{et al.}~\cite{yacheur2023drlrat} use a DRL-based communication mode selection strategy in a hybrid vehicular network using ITS-G5 and C-V2X, with the objective of improving reliability while limiting resource consumption. Their work focuses on RAT selection for vehicular services rather than evaluating 5G-WiFi mobility-driven handover frequency or offloading from cellular networks to WLANs. Allahham \textit{et al.}~\cite{allahham2022multiagent} formulate dynamic network selection and RAN resource allocation in heterogeneous multi-RAT networks as a deep multi-agent RL problem. Their formulation considers application QoS, energy, latency, cost, and resource allocation objectives rather than mobility-driven handover stability. De Santis \textit{et al.}~\cite{desantis2022satellite} propose a DQN-based controller for network selection and adaptive resource allocation in a heterogeneous 5G/satellite setting, with reported outcomes including connection acceptance, resource allocation, and load balancing. Habib \textit{et al.}~\cite{Habib2023trafficsteering,habib2023hierarchical} apply DRL and hierarchical RL to load-aware traffic steering between LTE and 5G NR while satisfying traffic QoS requirements. 
Learning based work that more directly considers heterogeneous mobility includes Rihani \textit{et al.}~\cite{rihani2018nnHOmultiRAT}, which develops a neural network-based mechanism for vertical handover decisions between WiFi and WiMAX. This work describes a neural network-based learning agent for vertical handover. Our comparison is therefore made at the level of target selection objective and evaluated setting, rather than claiming that its learning model is inferior or equivalent to the DQN formulation used here.

These studies address different, sometimes overlapping aspects of heterogeneous connectivity, including access selection, vertical handover, resource allocation, and traffic steering. These approaches do not capture the mobility-driven handover target selection problem addressed in this paper. Mobility-driven target selection additionally asks which specific candidate node should serve a moving UE at successive decision epochs. This introduces relevance for the number and direction of association changes alongside signal quality and delay. In Section~\ref{subsec_env_desc}, we evaluate the effectiveness of considering these additional parameters through a case study.

Finally, we observe that relatively few works examine how DRL-based handover frameworks can incorporate both network independent objectives and deployment specific operator preferences through reward design. A network independent criterion allows target selection without specifying separate thresholds for each RAT category. A deployment specific reward can instead encode relative operating characteristics and preferences, such as favoring a RAT type when configured QoS conditions are satisfied. This distinction is important because the same candidate selection problem can have different objectives in different heterogeneous deployments. Han \textit{et al.}~\cite{han2019a} use uplink throughput as a reward, Yajnanarayana \textit{et al.}~\cite{yajnanarayana2020a} pursue a beam selection objective, and Voigt \textit{et al.}~\cite{gu2024drl5g} incorporate signal quality and handover related outcomes in a PPO-based formulation. These formulations demonstrate the importance of reward design, but the reviewed studies do not explicitly compare a network independent formulation with a deployment specific formulation that incorporates RAT-specific constraints and operator preferences within the same handover framework. ReFlexMRS addresses this distinction by evaluating both a network independent reward formulation and a deployment specific formulation that incorporates RAT-specific thresholds and WiFi offloading preferences. 

As mentioned previously, SDN has unique architectural advantages, such as a managed global network view and programmability to manage QoS-selective multi-RAT handover. Recent SDN-based handover algorithms have explored handover among single-RAT networks~\cite{ciciouglu2021multi,rizkallah2018sdnverticalho} and go beyond simple RSSI-based handovers to consider QoS-based network selection for bandwidth, frequencies, and connection delays. Several works have shown that SDN can provide a unifying framework for multi-RAT operation~\cite{KHATURIA2021108412,Alfoudi2019seamless} and can facilitate the handover process by providing coordinated control~\cite{liyanage2017sdnoffloading}. However, these related techniques do not address the same combination of adaptive, multi-criteria, mobility-driven target selection considered here. In our previous work~\cite{mangipudi2024mcdm}, we examined multi-RAT handover using Multiple Criteria Decision Making (MCDM) with RSSI and delay as decision criteria. The present study instead examines learned policies with combined criterion and RAT-aware reward formulations, including a preference designed to reduce reliance on higher power access nodes when alternative QoS conditions are satisfied.

\begin{table*}[t]
\centering
\caption{Comparison of representative handover and multi-RAT learning approaches.}
\label{tab:related_work_comparison}
\footnotesize
\setlength{\tabcolsep}{2.5pt}
\setlength{\arrayrulewidth}{0.5pt}
\renewcommand{\arraystretch}{1.15}

\begin{tabular}{|
>{\raggedright\arraybackslash}p{0.155\textwidth}|
>{\raggedright\arraybackslash}p{0.195\textwidth}|
>{\raggedright\arraybackslash}p{0.205\textwidth}|
>{\raggedright\arraybackslash}p{0.135\textwidth}|
>{\raggedright\arraybackslash}p{0.24\textwidth}|}
\hline

\textbf{Approaches}
& \textbf{Key Ideas}
& \textbf{Assumptions / Scope}
& \textbf{Adaptivity}
& \textbf{Strengths and Limitations} \\
\hline

\textbf{MCDM/SDN Handover}~\cite{ciciouglu2021multi,rizkallah2018sdnverticalho,mangipudi2024mcdm}
& Centralized SDN handover and QoS-aware network selection; MCDM combines RSSI and delay for multi-RAT target selection.
& Includes single-RAT SDN schemes and a multi-RAT MCDM scheme. MCDM parameters must be customized for the network configuration and propagation environment.
& Rule or optimization-based, using predefined weights or thresholds rather than an adaptively learned policy.
& Uses centralized control and criteria beyond RSSI. Parameter tuning is deployment dependent, and the continued influence of RSSI can favor higher power nodes. \\
\hline

\textbf{Single-RAT RL/DRL Handover}~\cite{han2019a,tan2022a,mollel2020doubledrl,yajnanarayana2020a,gu2024drl5g}
& RL/DRL policies learn handover, AP selection, or beam selection decisions from objectives related to signal strength and throughput; some methods also seek to reduce frequent handovers.
& Standalone WLAN or cellular/mmWave environments in which candidate access nodes belong to one RAT; objectives often emphasize signal/link quality, throughput, or handover stability.
& Learning-based policies refined through interaction or experience; some reward designs also use signal strength conditions and thresholds.
& Demonstrates adaptive handover optimization under dynamic mobility and channel conditions. Does not address mobility-driven selection across heterogeneous RATs while jointly considering RSSI, delay, stability, and RAT-aware offloading. \\
\hline

\textbf{Multi-RAT RL Network Selection / Traffic Steering}~\cite{yacheur2023drlrat,allahham2022multiagent,desantis2022satellite,Habib2023trafficsteering,habib2023hierarchical}
& DRL, multi-agent RL, and hierarchical RL support communication mode or network selection, joint resource allocation, and load-aware traffic steering.
& Evaluated in hybrid ITS-G5/C-V2X, heterogeneous multi-RAT, 5G/satellite, and LTE/5G NR settings; the primary focus is association, resource allocation, or traffic steering.
& Learning-based policies adapt selection, allocation, or steering to QoS, load, reliability, and resource conditions.
& Addresses heterogeneous access and multiple service or resource objectives. Does not evaluate the same mobility-driven 5G-WiFi handover frequency, handover direction, and WiFi offloading problem. \\
\hline

\textbf{Learning-Based Vertical Handover}~\cite{rihani2018nnHOmultiRAT}
& A neural network based mechanism selects vertical handovers between WiFi and WiMAX.
& Heterogeneous WiFi-WiMAX mobility rather than the 5G-WiFi setting considered here.
& Learning-based neural network decision mechanism.
& Addresses WiFi--WiMAX vertical handover using a neural network learning agent; different setting and learning description from the evaluated DQN-based 5G-WiFi target selection task. \\
\hline

\textbf{ReFlexMRS (This Work)}
& {Configurable DQN-based, multi-criteria handover target selection across candidate nodes from heterogeneous RATs; network independent and deployment specific reward formulations allow RAT-aware offloading.}


& General decision formulation is multi candidate rather than pairwise; an empirical validation is provided using a 5G-WiFi case study with offline training and a gNB-like emulated node, not native NR procedures.


& DQN policy learned offline; the state and reward design can be configured for network independent criteria or deployment specific RAT preferences.
& Case study evaluates delay, throughput, total handovers, and WiFi-to-5G handovers within the ReFlexMRS framework. Current evidence is limited to the controlled emulation and does not include online closed-loop training or standardized 5G handover procedures. \\
\hline

\end{tabular}
\end{table*}

\subsection{Summary}

Table~\ref{tab:related_work_comparison} compares representative methods using the features most relevant to ReFlexMRS. A distinction is made between multi-RAT network selection or traffic steering and mobility-driven handover optimization. Prior studies address individual elements of the problem, including DRL-based RAT selection, multiple QoS requirements, and load-aware traffic steering. ReFlexMRS focuses on their combined application to 5G-WiFi handover target selection, configurable reward design, RAT-aware offloading, and handover stability evaluation.

Existing research demonstrates that RL can support single-RAT handover optimization as well as multi-RAT network selection, resource allocation, and traffic steering. However, these studies generally address individual portions of the broader mobility management problem: single-RAT handover methods commonly optimize signal strength or throughput, while heterogeneous multi-RAT methods primarily address access association, resource allocation, or traffic steering rather than mobility-driven handover target selection. To address these gaps, we propose a DRL-based approach that optimizes handover decisions across heterogeneous RATs using multiple metrics beyond RSSI. Our contributions are summarized as follows.

\begin{itemize}
    \item We implement DRL-based multi-metric optimization in the reward function for QoS-aware handover target selection.
    \item We develop a unified DRL-driven framework, ReFlexMRS, for handover target selection among heterogeneous RATs while accounting for RAT characteristics.
    \item We develop network independent and deployment specific state/reward formulations to demonstrate how RAT-specific information and operator preferences can be incorporated into the same framework.
\item We present a 5G-WiFi case study of ReFlexMRS and evaluate QoS and mobility management tradeoffs such as link delay, throughput, total handovers, and WiFi-to-5G handovers relative to conventional RSSI-based selection.
\end{itemize}

\begin{figure}[t!]
\centerline{\includegraphics[width=0.8\columnwidth]{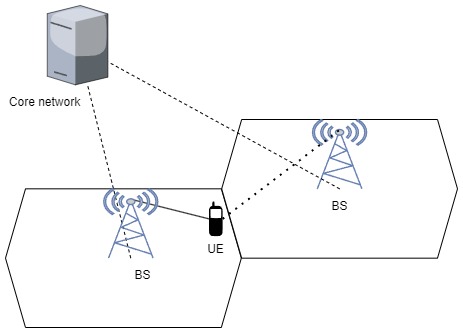}}
\caption{Simple cellular network topology during a handover.}
\label{simplehandover}
\end{figure}

\section{Background}\label{sec_background}
Handovers are a fundamental mechanism in cellular networks, ensuring continuous and reliable connectivity as User Equipment (UE) moves throughout the network. During handover, a UE transitions its connection, or point of attachment to the network, from one node (access point, gateway, or base station) to another or from one RF channel to another, as illustrated in Figure \ref{simplehandover}. Effective handover mechanisms are essential for preventing dropped calls, maintaining data session continuity, and optimizing the overall user experience. Handovers in cellular networks can occur for various reasons, including user mobility across nodes, switching a user to a different channel to avoid interference, or for system performance reasons, such as balancing the load by transferring users between overlapping cells when the capacity of one cell is nearly reached. The handover process generally involves three stages: the handover decision, which entails collecting and analyzing data to detect the need for a handover and triggering it; radio link transfer, which involves determining paths and routing; and channel assignment, which involves selecting the appropriate radio frequency channel~\cite{mcnairvertho2004}. This paper primarily focuses on the handover decision aspect. Accordingly, it does not attempt to model the full standardized signaling, interruption timing, or failure procedures of complete 3GPP handover stacks.

\begin{figure}[t!]
\centerline{\includegraphics[width=0.8\columnwidth]{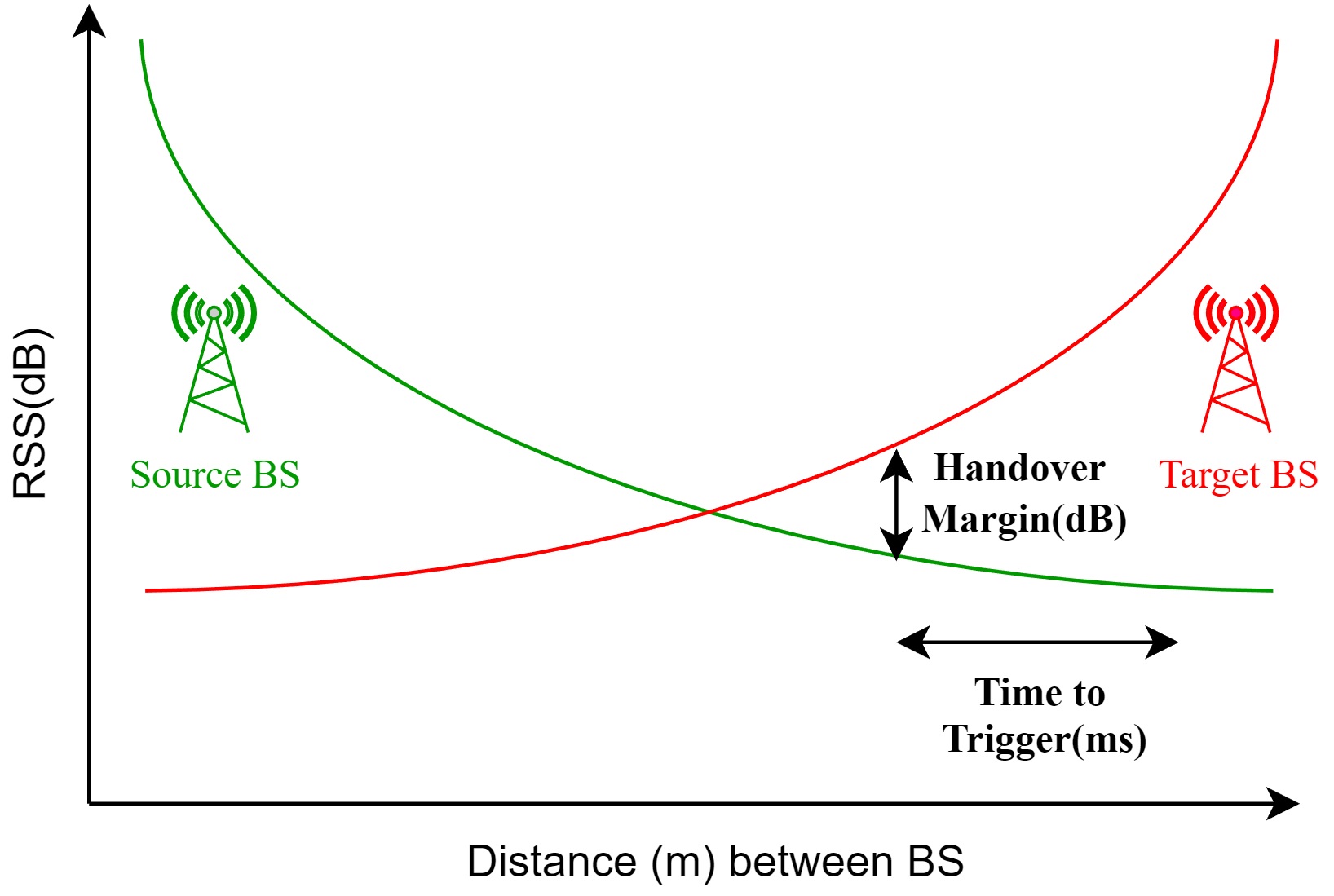}}
\caption{Hysteresis-based general handover}
\label{hysteresis}
\end{figure}

The node currently connected to the UE is termed the source node, while the node the UE is moving towards is referred to as the target node. Traditionally, a handover decision is initiated by the source node when the Received Signal Strength (RSS) of the UE at the source node drops below the RSS of the UE at the target node. Different generations of cellular communication standards specify various methods for handover initiation. An example of a handover triggering event is depicted in Fig. \ref{hysteresis}. The handover margin is a range, measured in dB, beyond which the UE's connection with the source node is lost due to insufficient received signal strength, RSS. Thus, the handover is triggered once the RSS at the source node decreases to dip below the Handover Margin. There can be other parameters besides the RSS that improve handover performance. The Time-to-Trigger (TTT) parameter determines how long the source node must wait after the Handover Margin is crossed before initiating the handover. This can be employed to ensure that the handover initiation decision does not cause the UE to ping pong between nodes.

Handovers in cellular networks can also be categorized into horizontal and vertical handovers based on the homogeneity or heterogeneity of the RATs involved in the handovers. A horizontal handover occurs between nodes using the same radio access network (RAN), and a vertical handover is when a UE must handover to a different RANs and RATs. Vertical handovers have become increasingly common due to the coexistence of legacy technologies such as LTE and 5G within newer B5G networks. Moreover, vertical handovers can frequently occur between cellular and WiFi as well as between different sizes of network nodes such as micro cells and macro cells, within the same RAT. To distinguish between them, we refer to vertical handovers to a different radio access technology as a multi-RAT handover. To perform handover in a multi-RAT environment, it is assumed that user equipment (UEs) has the capability to adapt to operate in multiple RATs. We note that this capability is being developed in RF, but is out of the scope of this paper \cite{wwangcoexist2023}.

In a multi-RAT environment, the handover decision process is particularly critical for leveraging the pros and cons of different networks in selecting a target RAN and maintaining seamless communication across different network types, such as 5G, WiFi, B5G and soon 6G technologies. Emerging research favors the use of software-defined networking (SDN) to segregate the control and data planes of each RAT, and connect each RAT through the control plane, addressing the implementation constraints of multi-RAT networks \cite{KHATURIA2021108412,liyanage2017sdnoffloading, alotaibi2018hierarchical, wang2016convergence}.

\subsection{Software-Defined Networking (SDN) }\label{subsec_sdn}

The OpenFlow protocol \cite{openflow} was the pioneer protocol developed for SDN and is maintained by the Open Networking Foundation (ONF) \cite{onf}. Instructions issued from the SDN Controller to the switch are termed flows, which encompass packet match fields, flow priority indicators, various counters, packet processing directives, flow timeouts, and cookies. These flows are organized within structures known as flow tables. In switches enabled with OpenFlow, flow tables and a group table oversee packet lookup and forwarding. OpenFlow channels, utilizing the Southbound Interface, link the SDN Controller to the data plane. The SDN Controller, through the OpenFlow protocol, possesses the capability to add, update, and remove flow entries in the flow tables, either reactively or proactively. This centralized controller formulates routing decisions and instructs the user plane through the flow tables embedded in OpenFlow switches.

The application of SDN in cellular networks is well documented in literature, with integrations into LTE and 5G NR architectures aimed at refining handover control, optimizing mobile routing, and creating efficient routing policies to reduce control message exchanges \cite{prados2016handover,rizkallah2018sdnverticalho,abdulghaffar2021modeling}. Nonetheless, these methodologies have primarily focused on single-RAT deployments. This study leverages SDN's control and data plane separation alongside reinforcement learning to realize an intelligent multi-RAT handover. The programmable centralized control provided by SDN facilitates network configuration and accommodates advanced decision-making algorithms, making it a suitable platform for implementing Machine Learning (ML) techniques \cite{mangipudi2023survey}.

\begin{figure}[t!]
\centerline{\includegraphics[width=0.8\columnwidth]{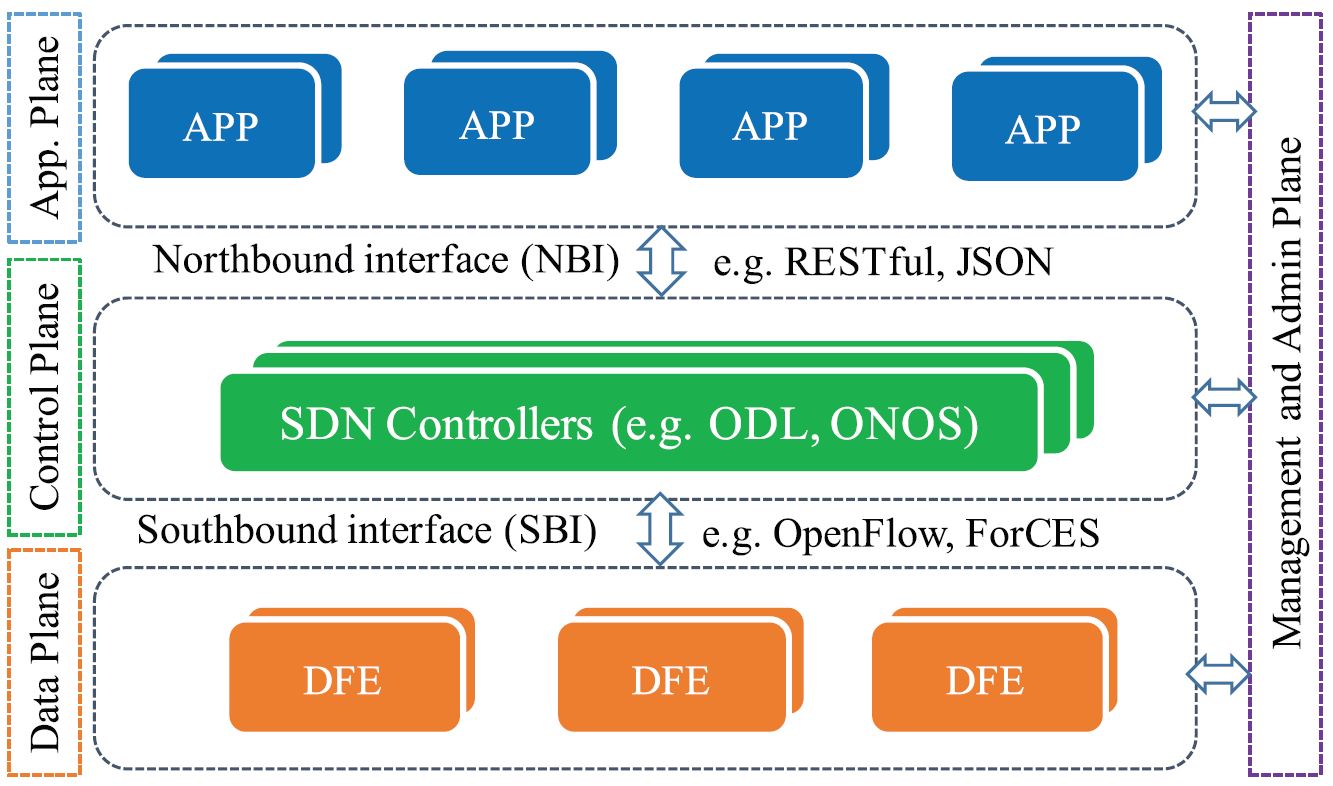}}
\caption{A high-level SDN architecture \cite{nguyen2017sdnnfvltesurvey}.}
\label{SDNarch}
\end{figure}

\subsection{Reinforcement Learning and Deep Q Networks}
\label{subsec_bg_rl_dqn}

Reinforcement Learning (RL) is a framework in machine learning where an agent learns to make decisions by interacting with an environment to maximize cumulative rewards over time. At each time step, the agent observes a state $s \in \mathcal{S}$, chooses an action $a \in \mathcal{A}$, and receives a scalar reward $r \in \mathbb{R}$ that reflects the immediate benefit of taking that action. The environment is commonly modeled as a Markov Decision Process (MDP) \cite{sutton2018a}, which is defined by the tuple $(\mathcal{S}, \mathcal{A}, P, R, \gamma)$, where  $P(s'|s,a)$ is the transition probability to the next state $s'$ given current state $s$ and action $a$, $R(s,a)$ is the reward function, and $\gamma \in [0,1)$ is the discount factor that weighs future rewards relative to immediate ones.

A key objective in RL is to learn an optimal policy $\pi(a|s)$ that maximizes the expected return defined by
\begin{equation}
    G_t = \sum_{k=0}^{\infty} \gamma^k r_{t+k}.
\end{equation}
Central to many RL algorithms is the action-value function or Q-function, which measures the expected return when starting from state $s$, taking action $a$, and thereafter following policy $\pi$:
\begin{equation}
    Q^{\pi}(s,a) = \mathbb{E}_{\pi}\left[G_t \mid s_t = s, a_t = a\right].
\end{equation}
The optimal Q-function, denoted $Q^*(s,a)$, satisfies the Bellman optimality equation \cite{jang2019qlearning}:
\begin{equation}\label{bellman}
    Q^*(s,a) = \mathbb{E}_{s'} \left[r + \gamma \max_{a'} Q^*(s',a') \mid s, a\right].
\end{equation}

While traditional Q-learning uses a tabular method to represent $Q(s,a)$, this approach becomes infeasible for large or continuous state spaces. This is due to the curse of dimensionality, where the number of possible state-action pairs grows exponentially with the number of features, making it impractical to store or update a complete table. Deep Q Networks (DQNs) \cite{mnih2015Human} address this challenge by using a deep neural network with parameters $\theta$ to approximate the Q-function, denoted $Q(s,a;\theta)$. The DQN is trained to minimize the mean squared error between its current estimate and a target value. Specifically, the loss function is given by
\begin{equation}\label{loss}
    L(\theta) = \mathbb{E}_{(s,a,r,s') \sim D} \left[ \left( y - Q(s,a;\theta) \right)^2 \right],
\end{equation}
where the target $y$ is computed as
\begin{equation}\label{target}
    y = r + \gamma \max_{a'} Q(s',a';\theta^-)
\end{equation}

\noindent and $\theta^-$ represents the parameters of a target network that is periodically updated with the latest parameters $\theta$ from the local network. This decoupling is crucial because it prevents the target values from shifting too rapidly, which can otherwise lead to instability or divergence during training. In tandem with experience replay where transitions $(s, a, r, s')$  are uniformly sampled from a replay buffer $D$. This mechanism breaks the temporal correlations inherent in sequential data. As a result, it not only stabilizes the learning process by providing a more consistent target for the Q-value updates, but also enhances data efficiency by allowing the model to learn repeatedly from past experiences.

In this work, we use an SDN-based architecture for multi-RAT network control together with a Deep Reinforcement Learning (DRL)-based multi-criteria handover mechanism. The multi-criteria handover target selection problem is formulated as an MDP with a finite action space and a continuous-valued measurement representation, and a Deep Q-Network is trained to learn a target selection policy. The model additionally employs convolutional layers to process correlations among measurements collected from users and candidate access nodes, enabling the agent to learn the handover target selection policy.

\begingroup
\section{ReFlexMRS: General Framework and Methodology}\label{sec_methodology}

We propose ReFlexMRS as a configurable deep reinforcement learning (DRL) framework for multi-criteria handover target selection across heterogeneous radio access networks. The framework uses measurements from candidate access nodes to learn a selection policy according to the desired QoS objectives and network operating preferences. By defining the candidate nodes, decision criteria, and reward rules separately, the same framework can be applied to different multi-RAT configurations. In this study, we train the agent offline using collected network data and evaluate the learned policy in an SDN-enabled 5G-WiFi case study, described in Section~\ref{subsec_env_desc}.

\subsection{target selection State and Action Formulation}\label{subsec_mdp}

We consider a UE that can select among $N$ candidate access nodes, indexed by $\mathcal{A}=\{0,1,\ldots,N-1\}$. At each decision step $t$, we collect $K$ measurements for each candidate node $i\in\mathcal{A}$, represented by
\begin{equation}\label{eq:general_measurements}
\mathbf{x}_{i,t}=[x_{i,t}^{(1)},x_{i,t}^{(2)},\ldots,x_{i,t}^{(K)}]^{\mathsf T}.
\end{equation}

These measurements form the state representation used for handover target selection. We consider two ways to represent them: (i) a \emph{combined feature representation}, denoted by $\mathbf{s}_t^{(1)}$, and (ii) a \emph{separate feature representation}, denoted by $\mathbf{s}_t^{(2)}$. The two representations are evaluated as DRL1 and DRL2, respectively, in Section~\ref{subsec_env_desc}. In the combined feature representation, a function $g(\cdot)$ combines the $K$ measurements of each candidate into a single feature:
\begin{equation}\label{eq:general_combined_state}
\mathbf{s}_t^{(1)}=[g(\mathbf{x}_{0,t}),\ldots,g(\mathbf{x}_{N-1,t})]^{\mathsf T}\in\mathbb{R}^{N}.
\end{equation}
Alternatively, the separate feature representation retains each measurement as an individual feature, with the values grouped by criterion across the candidate nodes:
\begin{equation}\label{eq:general_separate_state}
\begin{split}
\mathbf{s}_t^{(2)}=[&x_{0,t}^{(1)},\ldots,x_{N-1,t}^{(1)},\ldots,\\
&x_{0,t}^{(K)},\ldots,x_{N-1,t}^{(K)}]^{\mathsf T}\in\mathbb{R}^{NK}.
\end{split}
\end{equation}
The resulting DQN input has $N$ features for the combined representation and $NK$ features for the separate representation. The measurements can be normalized or otherwise preprocessed according to the selected criteria. We describe the preprocessing and combination rule used in our case study in Section~\ref{subsec_case_preprocessing}.

At each decision step, the agent chooses an action $a_t\in\mathcal{A}$, where $a_t=i$ represents selection of candidate node $i$ as the handover target. Both feature representations therefore share the same $N$ possible actions. The agent receives continuous-valued measurement features, while the action space remains finite. The next state $\mathbf{s}_{t+1}$ represents the subsequent network observation used for DQN training.

\subsection{Configurable Reward Formulation}\label{subsec_general_reward}

The reward function guides the agent toward selecting candidate nodes that satisfy the chosen optimization objectives. Depending on the deployment, these objectives may involve a combined QoS measure, improvements in individual measurements, or preferences for particular categories of access nodes. We express the reward for the selected target at decision step $t$ as
\begin{equation}\label{eq:general_reward}
r_t=R_j(\mathbf{s}_t^{(j)},a_t,c_t),
\end{equation}
where $R_j$ denotes the configured reward function and $c_t$ identifies the access node currently serving the UE, obtained from its association information. The serving node index is used when the reward compares a candidate target with the current connection; it is not an additional DQN input feature. A reward based only on the candidate's combined QoS, for example, does not require $c_t$. In the 5G-WiFi case study, we evaluate two reward functions, $R_1$ and $R_2$, corresponding to the two feature representations. The reward and the next observation $\mathbf{s}_{t+1}$ are stored with $\mathbf{s}_t$ and $a_t$ for offline DQN training.

\subsection{DQN Architecture and Offline Policy Learning}\label{subsec_DQNarch}
\par\endgroup

\begin{figure*}[ht!]
\centerline{\includegraphics[width=0.8\textwidth]{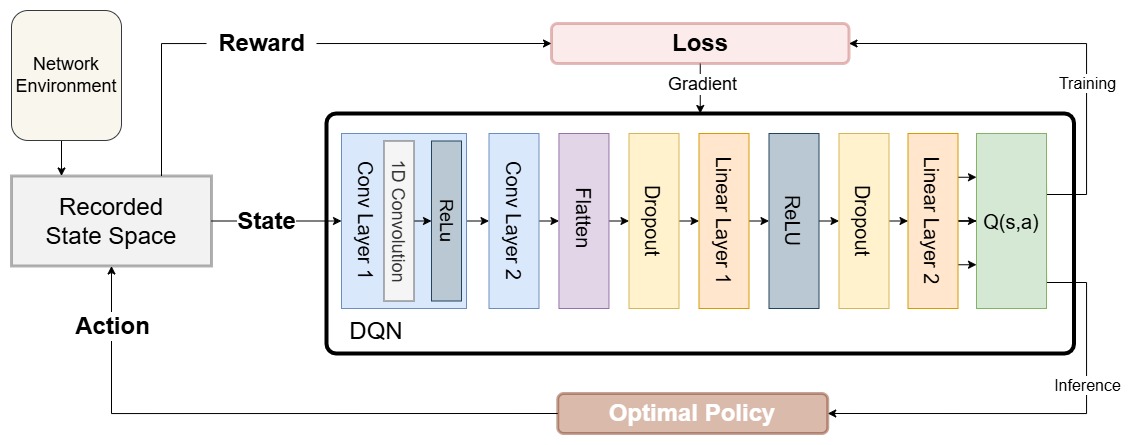}}
\caption{Block Diagram of the DQN-based handover agent}
\label{blockdiagram}
\end{figure*}

\begingroup
Fig.~\ref{blockdiagram} shows the DQN-based handover agent used in ReFlexMRS. The network approximates the action-value function $Q(\mathbf{s}_t,a;\theta)$ and estimates a Q-value for each of the $N$ candidate targets. It consists of two one-dimensional convolutional layers followed by two fully connected layers. We use the convolutional layers to extract local dependencies among network measurement features, which may contain information about signal quality, relative proximity, propagation, and other network conditions. The fully connected layers use the extracted features to estimate the action values, while dropout is applied during training to reduce overfitting.

To train the DQN offline, we first collect measurements from the network environment. At each training step, the agent selects a candidate target using the current observation, evaluates the action using the configured reward function, and proceeds to the next stored observation. The resulting experiences $(\mathbf{s}_t,a_t,r_t,\mathbf{s}_{t+1})$ are added to replay memory and sampled to train the local network using the Bellman target and loss described in Section~\ref{subsec_bg_rl_dqn}. We also maintain a target network with the same architecture, whose weights are periodically updated from the local network to improve training stability. After training, the learned network weights are used to select handover targets from newly collected measurements.

During offline training, we employ softmax action selection with temperature $\tau$ to balance exploration and exploitation. For a given observation $\mathbf{s}_t$, the probability of selecting candidate $a$ is
\begin{equation}\label{softmax_with_temp}
P(a\mid\mathbf{s}_t)=\frac{\exp\!\big(Q(\mathbf{s}_t,a;\theta)/\tau\big)}{\sum_{a'\in\mathcal{A}}\exp\!\big(Q(\mathbf{s}_t,a';\theta)/\tau\big)}.
\end{equation}
A higher temperature results in more uniform action probabilities and greater exploration, whereas a lower temperature favors actions with higher estimated Q-values. During inference, the trained agent selects the target with the highest estimated Q-value:
\begin{equation}\label{eq:learned_policy}
\hat a_t=\arg\max_{a\in\mathcal{A}}Q(\mathbf{s}_t,a;\theta).
\end{equation}
These Q-values define the learned handover target selection policy, which selects actions based on the estimated cumulative rewards associated with the observed network conditions. In the following section, we describe how the state, reward, and DQN training formulations are implemented in the 5G-WiFi case study.
\par\endgroup

\section{Case Study and Example Implementation}\label{subsec_env_desc}
\begingroup
In this section, we evaluate the ReFlexMRS framework in a controlled 5G-WiFi environment using an SDN-enabled multi-RAT network. The SDN controller supports network measurement collection and the application of handover target selection decisions. We first describe the emulated network topology, followed by the state and reward formulations and the DQN training configuration used for DRL1 and DRL2.
\par\endgroup

\begin{table}[t!]
\centering
    \caption{Mininet-WiFi parameters for multi-RAT network simulation.} \label{network_setup} 
    \resizebox{0.45\textwidth}{!}{%
    \begin{tabular}{lll} 
    \toprule
    Parameter & WiFi AP & 5G gNB-like emulated node \\    
    \midrule
    Number & 13 & 1 \\    
    \midrule
    Frequency & 5.18-5.825GHz & 2.412GHz\\
    \midrule
    TxPower & 2dBm & 16dBm\\
    \midrule
    Range & 60m & 150m\\
    \midrule
    Mode & ax5 & g\\
    \midrule
    Propagation Model & LogNormal & LogNormal \\
    \midrule
    Pathloss exponent & 3.5 & 3.5 \\ 
\bottomrule
\end{tabular}}
\vspace{2pt}

\parbox{0.95\columnwidth}{\footnotesize \textit{Note:} In Mininet-WiFi, the mode field specifies the underlying IEEE 802.11 wireless mode used by the emulator. The value ``g'' is used as the available Mininet-WiFi wireless mode for the customized 5G gNB-like node, while the gNB-like behavior is represented through customized transmit power, coverage range, and propagation parameters rather than native 3GPP-compliant 5G NR MAC/PHY support.}
\end{table}

\begin{figure}[t!]
\centerline{\includegraphics[width=\columnwidth]{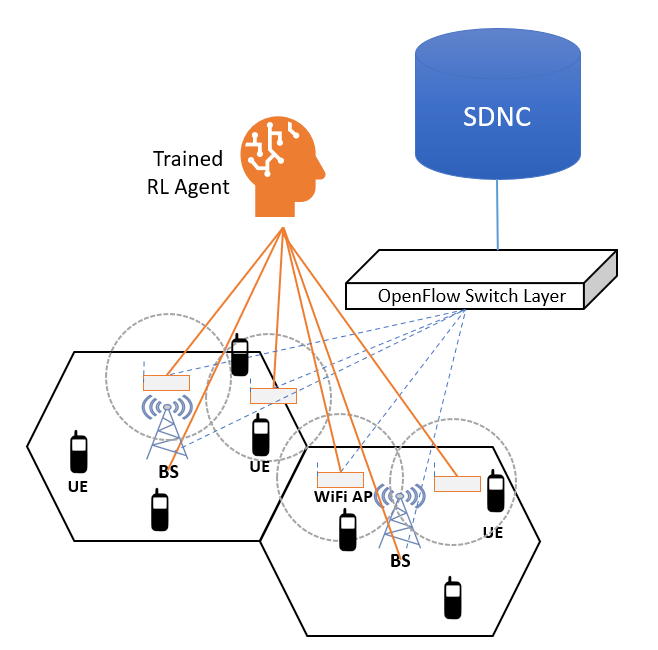}}
\caption{SDN-enabled multi-RAT evaluation architecture.}
\label{ENVarchitecture}
\end{figure}
Fig.~\ref{ENVarchitecture} illustrates the SDN-enabled network used in the 5G-WiFi case study. We adopt SDN because its centralized network view and programmability support measurement collection, association control, and packet forwarding across the emulated access networks. The gNB-like emulated node and WiFi access points (APs) are connected through OpenFlow-enabled switches managed by a centralized SDN controller. When an incoming packet does not match an existing flow rule, the switch sends its information to the controller, which determines the forwarding path and installs the appropriate rule. Subsequent matching packets are then forwarded through the selected access path. This arrangement allows the controller to support handover target selection decisions in the emulated network.

We assume that user equipment (UE) can operate across the RATs considered in the emulation. The adaptive RF and medium access capabilities required for such operation are under active investigation~\cite{wwangcoexist2023} and are outside the scope of this study. For each UE, RSSI and average link delay are collected periodically from the access nodes within its coverage range. These measurements provide the observations used to train the DRL agent and to evaluate candidate handover targets. Previous work~\cite{ciciouglu2021multi} has also demonstrated that handover optimization and mobility management can be implemented at an SDN controller through modifications to OpenFlow messages. 
Following offline training, the DRL agent is integrated with the emulated network to select target nodes during inference, as illustrated in Fig.~\ref{ENVarchitecture}. Our evaluation focuses on these target selection decisions, rather than the complete lower-layer handover procedures.

\subsection{Emulated Network Architecture and Topology}
We implemented the case study network using the Mininet-WiFi emulator for wireless SDN research, with OpenFlow-enabled switches and the default POX controller. Because Mininet-WiFi does not natively support 5G NR base stations, we implement a gNB-like node through custom modifications and parameter-matched emulation. Table \ref{network_setup} lists the parameters used for each RAT. For the gNB-like node, the operating frequency is set to 2.412~GHz, and the transmit power is determined using the 3GPP Non-Line-of-Sight Urban Macro model for 5G NR \cite{5gnrchannelmodel}. The calculated transmit power and range values are mapped to corresponding values available within the Mininet-WiFi urban mobility environment \cite{goldsmith2005wireless}. The Log-Normal propagation model is selected with a path loss exponent of 3.5. The customized transmit power, coverage range, propagation characteristics, and available Mininet-WiFi wireless configuration are used to represent gNB-like behavior. Accordingly, this setup should be interpreted as a controlled emulation of cross-RAT handover decision-making rather than a native 3GPP-compliant 5G NR MAC/PHY implementation or a reproduction of full deployment realism. For the WiFi APs, the operating frequency is set to 5~GHz in accordance with the 802.11ax standard.


\begingroup
The experimental topology consists of one gNB-like emulated node and 13 WiFi APs, with the WiFi APs evenly distributed within the coverage area of the gNB-like node. We randomly distribute 64 UEs across the topology and use the ``random direction'' mobility model in Mininet-WiFi. The UEs move at 1.44 kilometers per hour (approximately 0.4 meters per second), representing low-speed mobility in a densely populated urban environment.

In the general target selection formulation of Section~\ref{sec_methodology}, this topology corresponds to $N=14$ candidate access nodes, with $\mathcal{A}=\{0,\ldots,13\}$. We assign index~0 to the gNB-like node and indices~1-13 to the WiFi APs. We use $K=2$ decision criteria for each candidate node: RSSI and average link delay. Applying Eqs.~\eqref{eq:general_combined_state} and~\eqref{eq:general_separate_state}, the combined feature representation therefore contains $N=14$ input features, whereas the separate feature representation contains $NK=28$ features. Both representations use the same set of 14 candidate actions. Their specific measurement and reward formulations are described below.
\par\endgroup

\begingroup
\subsection{Case Study State and Reward Instantiation}\label{subsec_case_mdp}

For each UE, we collect RSSI and average link delay from the access nodes within its coverage range. To maintain a fixed number of input features, we represent all 14 candidate nodes in each observation. For out-of-range nodes, we assign default values of $-150$~dB for RSSI and $1$~s for delay. These values are used for preprocessing rather than representing measured links; no separate action mask is applied. DRL1 and DRL2 use the same observations and candidate actions but differ in how the measurements are represented and rewarded. We describe these formulations using the general measurement, state, and reward definitions introduced in Section~\ref{sec_methodology}.

\subsubsection{Common QoS Preprocessing}\label{subsec_case_preprocessing}
For candidate node $i$ at decision step $t$, let $RSSI_{i,t}$ and $d_{i,t}$ denote the RSSI and average link delay, respectively. We normalize RSSI across the candidate nodes in each observation. The delay measurements are similarly normalized and then inverted, so that higher values of either resulting feature correspond to better QoS:
\begin{equation}\label{eq:rssi_normalization}
\widetilde{RSSI}_{i,t}=\frac{RSSI_{i,t}-RSSI_{\min,t}}{RSSI_{\max,t}-RSSI_{\min,t}},
\end{equation}
\begin{equation}\label{eq:delay_normalization}
\widetilde d_{i,t}=1-\frac{d_{i,t}-d_{\min,t}}{d_{\max,t}-d_{\min,t}}.
\end{equation}
Here, the minima and maxima are calculated across the $N=14$ candidate nodes at decision step $t$, assuming a nonzero range for each measurement. Following preprocessing, the general measurement vector of Eq.~\eqref{eq:general_measurements} becomes $\mathbf{x}_{i,t}=[\widetilde{RSSI}_{i,t},\widetilde d_{i,t}]^{\mathsf T}$. Both DRL1 and DRL2 use these same normalized RSSI and inverted delay values; the difference is whether the two criteria are combined or retained separately as state features.

\subsubsection{DRL1: Combined QoS Features and Reward}\label{subsec_case_drl1}
For DRL1, we apply the combined feature representation in Eq.~\eqref{eq:general_combined_state} by defining $g(\mathbf{x}_{i,t})$ as the product of the normalized RSSI and inverted delay features of candidate node $i$:
\begin{equation}\label{eq:case_combined_qos}
QoS_{i,t}=\widetilde{RSSI}_{i,t}\,\widetilde d_{i,t}.
\end{equation}
Thus, $g(\mathbf{x}_{i,t})=QoS_{i,t}$ and the state is $\mathbf{s}_t^{(1)}=[QoS_{0,t},\ldots,QoS_{13,t}]^{\mathsf T}$, containing one QoS feature per candidate node. This formulation does not require RAT-specific thresholds. Instead, we reward the selection of the candidate with the highest combined QoS in the current observation, denoted by $QoS_{\max,t}=\max_{i\in\mathcal{A}}QoS_{i,t}$. For a selected target $a_t$, the reward $R_1$ of the general formulation in Eq.~\eqref{eq:general_reward} is defined as
\begin{equation}\label{eq:reward_function_1}
R_1=\begin{cases}
+1, & \text{if }QoS_{a_t,t}=QoS_{\max,t},\\
-1, & \text{otherwise.}
\end{cases}
\end{equation}

\subsubsection{DRL2: Separate QoS Features and RAT-Aware Reward}\label{subsec_case_drl2}
For DRL2, we apply the separate feature representation of Eq.~\eqref{eq:general_separate_state}. Rather than combining RSSI and inverted delay into one QoS value, we retain the two measurements separately for all candidate nodes. The resulting 28-feature state is
\begin{equation}\label{eq:case_separate_state}
\begin{split}
\mathbf{s}_t^{(2)}=[&\widetilde{RSSI}_{0,t},\ldots,\widetilde{RSSI}_{13,t},\\
&\widetilde d_{0,t},\ldots,\widetilde d_{13,t}]^{\mathsf T}.
\end{split}
\end{equation}
The reward $R_2$ instantiates Eq.~\eqref{eq:general_reward} with separate RSSI and delay improvement conditions and RAT-specific preferences.
\par\endgroup

We selected the thresholds and reward levels in $R_2$ heuristically by examining the observed ranges and variability of the normalized measurements in the collected traces, rather than through a hyperparameter grid search. These values reflect the considered RAT characteristics and operator preferences.

The primary objective is to maximize RSSI while simultaneously minimizing link delay, thereby balancing strong signal quality and low latency. Since the reward function uses the inverted normalized delay feature, maximizing the delay feature is equivalent to improving delay performance. In addition to this main objective, we implement a RAT-aware, offloading-oriented handover preference. Because the gNB-like node operates at substantially higher transmit power than the WiFi APs, its RSSI can dominate target selection even when a WiFi AP provides adequate signal quality and delay. The reward function therefore includes a penalty for selecting the gNB-like node, represented by action index 0, based on its performance relative to the best alternative WiFi AP. Here, $alt\_max\_RSSI$ and $alt\_max\_delay$ denote the maximum RSSI and inverted delay feature, respectively, among the available WiFi APs. The RSSI margin of 0.15 was introduced to reduce the influence of the gNB's inherent transmit power advantage, while the inverted delay margin of $0.01$ provides a corresponding cross-RAT comparison for latency. If the gNB exceeds the best WiFi alternative by either margin but is not best overall in both metrics, it receives the moderate penalty $-0.25$. This discourages gNB selection based primarily on an advantage in a single criterion without excluding the gNB entirely. Conversely, selecting a WiFi AP while obtaining a sufficient RSSI improvement or delay reduction is assigned a positive reward. Overall, this reward structure enables the agent to balance signal strength, latency, and RAT selection preferences while encouraging WiFi offloading when its QoS improvement is sufficient.

\begingroup
To evaluate the reward, we use the UE's current serving node index $c_t$, obtained from its association information, and the selected target node index $a_t$. The ``before'' and ``after'' values are the normalized measurements of the serving node and candidate target, respectively, from the same observation:
\begin{align}
RSSI_{before} &= \widetilde{RSSI}_{c_t,t}, & RSSI_{after} &= \widetilde{RSSI}_{a_t,t},\label{eq:r2_rssi_mapping}\\
delay_{before} &= \widetilde d_{c_t,t}, & delay_{after} &= \widetilde d_{a_t,t}.\label{eq:r2_delay_mapping}
\end{align}
Thus, the reward compares the current connection with the prospective target rather than using measurements taken after a handover. In the following conditions, $RSSI_{max}$ and $delay_{max}$ denote the largest normalized RSSI and inverted delay values across the $N$ candidates at decision step $t$.
\par\endgroup

Improvements in RSSI and delay performance are defined as:
\begin{equation}
    RSSI_{imp}=RSSI_{after}-RSSI_{before} \label{eq:improvemet_eq}
    \end{equation}
\begin{equation}
    delay_{imp}=delay_{after}-delay_{before}
    \label{eq:delay_eq}
\end{equation}

\noindent and the reward function \(R_2\) is defined:
\begin{equation}
\resizebox{\columnwidth}{!}{$
      R_2 = \begin{cases}
+1, & \text{if } RSSI_{after} = RSSI_{max} \text{ and } \\
& delay_{after} = delay_{max}, \\[1mm]
-0.25, & \text{if } action = 0 \text{ and } \\[1mm]
  & \bigl[(RSSI_{after} - alt\_max\_RSSI) > 0.15 \text{ or } \\[1mm]
  & (delay_{after} - alt\_max\_delay) > 0.01\bigr], \\[1mm]
+0.75, & \text{if } action \neq 0 \text{ and } \\[1mm]
  & (delay_{imp} > 0.05 \text{ or } RSSI_{imp} > 0.1), \\[1mm]
+0.5, & \text{if } delay_{imp} > 0.05 \text{ and } RSSI_{imp} > 0.1, \\[1mm]
-1, & \text{otherwise.}
\end{cases}
$}
\label{eq:reward_function_2}
\end{equation}

The reward conditions in Eq.~\eqref{eq:reward_function_2} are evaluated sequentially in the order presented. A target that is best among all candidates in both RSSI and inverted delay receives the maximum reward of $+1$, regardless of its RAT. A WiFi selection, represented by $action \neq 0$, receives $+0.75$ when its normalized RSSI improves by more than 0.1 or its inverted delay feature improves by more than 0.05 relative to the UE's current connection. Because the delay feature is inverted, the latter condition represents a reduction in actual link delay. Thus, WiFi offloading is encouraged only when the selected AP provides a meaningful improvement in at least one QoS criterion. A remaining target selection that simultaneously improves RSSI and inverted delay by more than 0.1 and 0.05, respectively, receives $+0.5$. The gNB penalty of $-0.25$ is less severe than the default reward of $-1$, allowing the gNB to remain preferable when the available WiFi APs do not satisfy the improvement criteria. A selection that satisfies none of the specified QoS or RAT selection conditions receives $-1$.

\begingroup
\subsection{Training Configuration and Inference}\label{subsec_case_training}

We train separate DQN agents for DRL1 and DRL2 using the architecture described in Section~\ref{subsec_DQNarch} and their respective state and reward formulations. Each network estimates 14 Q-values, corresponding to the candidate handover targets defined by $\mathcal{A}$. Table~\ref{hyperparameters} summarizes the training settings for both agents, including input dimensions, convolutional filters, and target network update frequency.
\par\endgroup

\begin{table}[h]
\centering
    \caption{Training Hyperparameters} \label{hyperparameters} 
    \resizebox{0.85\columnwidth}{!}{%
    \begin{tabular}{c c} 
    \toprule
    Parameter & Value\\    
    \midrule
    Convolution layer filters & 16, 32\\
    \midrule
    Hidden Layer size & 64\\
    
    \midrule
    Input Layer Size & 14 (DRL1), 28 (DRL2)\\
    \midrule
    Output Layer Size & 14\\
    \midrule
    Dropout factor  & 0.5 \\
    \midrule 
    Discount Factor \(\gamma\) & 0.99 \\
    \midrule       
    Maximum timesteps & 1000 \\
    \midrule
    Episodes & 750 \\
    \midrule    
    Mini-Batch Size & 64\\
    \midrule    
    Target network 
    replacement frequency & 200\\
    \midrule    
    Optimizer & Adam \\
    \midrule
    Learning Rate \(\alpha\) & 0.001\\
    \midrule     
    Weight Decay & 0.001\\
    \midrule
    Replay Memory Size & 10,000\\
    \midrule
    Initial and Final temperature \(\tau\) & 1, 0.01\\ 
    \midrule
    Temperature \(\tau\) Decay rate & 0.995\\

\bottomrule
\end{tabular}}
\end{table}

We initialized the training hyperparameters using settings commonly employed in DQN implementations and refined them through a grid search over predefined values on the validation set. During training, the softmax temperature begins at 1.0 and is multiplied by 0.995 after each step, producing a geometric decay (e.g., $1.0$, $0.995$, $0.995^2$, $\ldots$) until the minimum value of 0.01 is reached. The hyperparameters in Table~\ref{hyperparameters} were fixed after validation and before final testing.

At decision time, the trained agent requires a single forward pass through the convolutional and fully connected layers. With fixed filter sizes and hidden layer dimensions, inference complexity scales approximately linearly with the number of input measurements. The resulting decision time overhead is therefore expected to be small relative to offline training.

\begingroup
After training, the learned network weights are used for handover target selection during inference. For each new observation, the DQN estimates the action values and selects the candidate with the highest estimated Q-value, as given in Eq.~\eqref{eq:learned_policy}.

The next section presents the numerical results for the two ReFlexMRS formulations evaluated in the SDN-enabled 5G-WiFi environment.
\par\endgroup

\section{Numerical Results}
\label{sec_results}

Experiments were performed to compare traditional RSSI-based handover with two new handover decision processes based on the proposed DRL agents approach. Three metric categories were selected to analyze the handover performance: (1) link delay difference, (2) handover counts, including total handovers and WiFi-to-5G handovers, and (3) average cumulative throughput.
For each metric, we report the average value and corresponding standard deviation obtained from seven simulation runs, each lasting 60 minutes. These results characterize framework feasibility and trade-off behavior in one controlled emulated scenario and should not be interpreted as broad generalization across deployments, mobility models, or full protocol stacks.

 \begin{figure}[t]
\centerline{\includegraphics[width=\columnwidth]{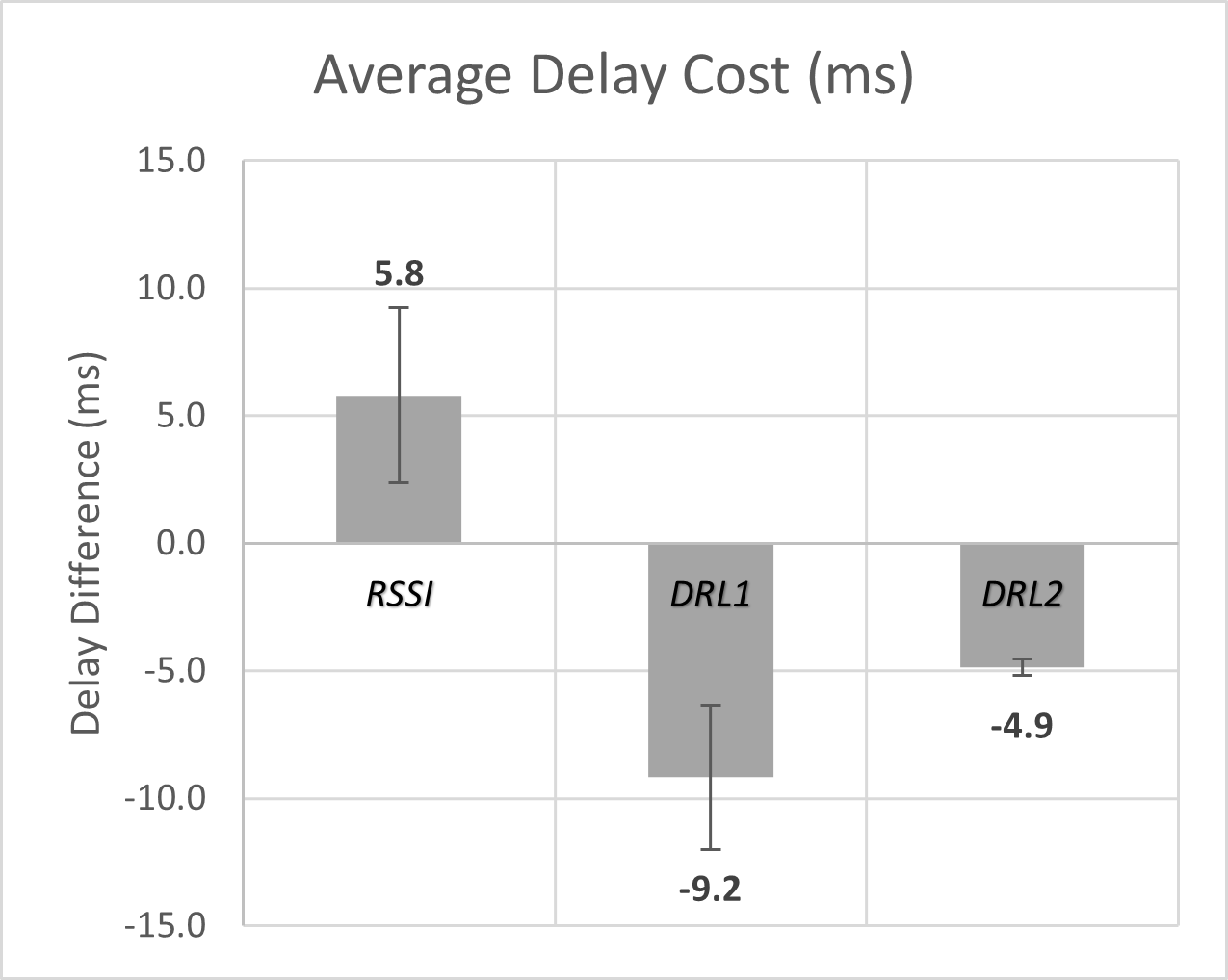}}
\caption{Delay Difference comparison}
\label{delay}
\end{figure}

\subsection{Difference in Link Delay}
\label{subsect:delayresults}
Fig.\ref{delay} presents a comparison of the difference in link delay for each model.  The mean round-trip time (RTT) is the time measured for a series of Internet Control Message Protocol (ICMP) Echo Request packets to be sent from a source host to a target host and back. The average link delay is calculated by taking the sum of the RTTs for all packets and dividing by the number of packets sent.This is used to provide a measure of the network latency between the source UE and target nodes. The link delay difference is the change in the average link delay experienced by the UE before and after the handover. It is calculated as the difference between the average delay experienced at the target node and the average delay experienced at the previous node. A negative value indicates that the handover delay at the target node was reduced, i.e., maintains a lower link delay for the UE's connection. A positive value means the selected target connection point causes increases the average link delay for the UE. As seen in Fig.\ref{delay}, using an RSSI-based handover decision in a multi-RAT scenario resulted in an increased average link delay after the handover occurred. On the other hand, the new DRL agents technique results in significant improvements in average link delay. DRL1 achieves a 258.6\% relative improvement in link delay benefit, while DRL2 achieves a 184.5\% relative improvement. This delay improvement occurs while ReFlexMRS retains most of the baseline throughput, with DRL1 maintaining approximately 90\% and DRL2 approximately 95\% of the RSSI-based approach's throughput. 

The standard deviations for the baseline RSSI, DRL1, and DRL2 handover methods are 6.85, 5.68, and 0.65, respectively. The traditional RSSI-based decision did not encourage or discourage delay improvements, and resulted in greater variation for the delay differences between source and target nodes. DRL1 exhibited a larger standard deviation, which is consistent with the less specific delay feature tuning of the DRL1/$R1$ combined feature reward function. The tight standard deviation for DRL2 shows the consistency achieved by the DRL2/$R2$ methodology, which employs a multifaceted approach with specific thresholds and prioritization for each of RSSI and delay, aiming for a balanced optimization across different performance metrics. 

\begin{figure}[t]
\centerline{\includegraphics[width=\columnwidth]{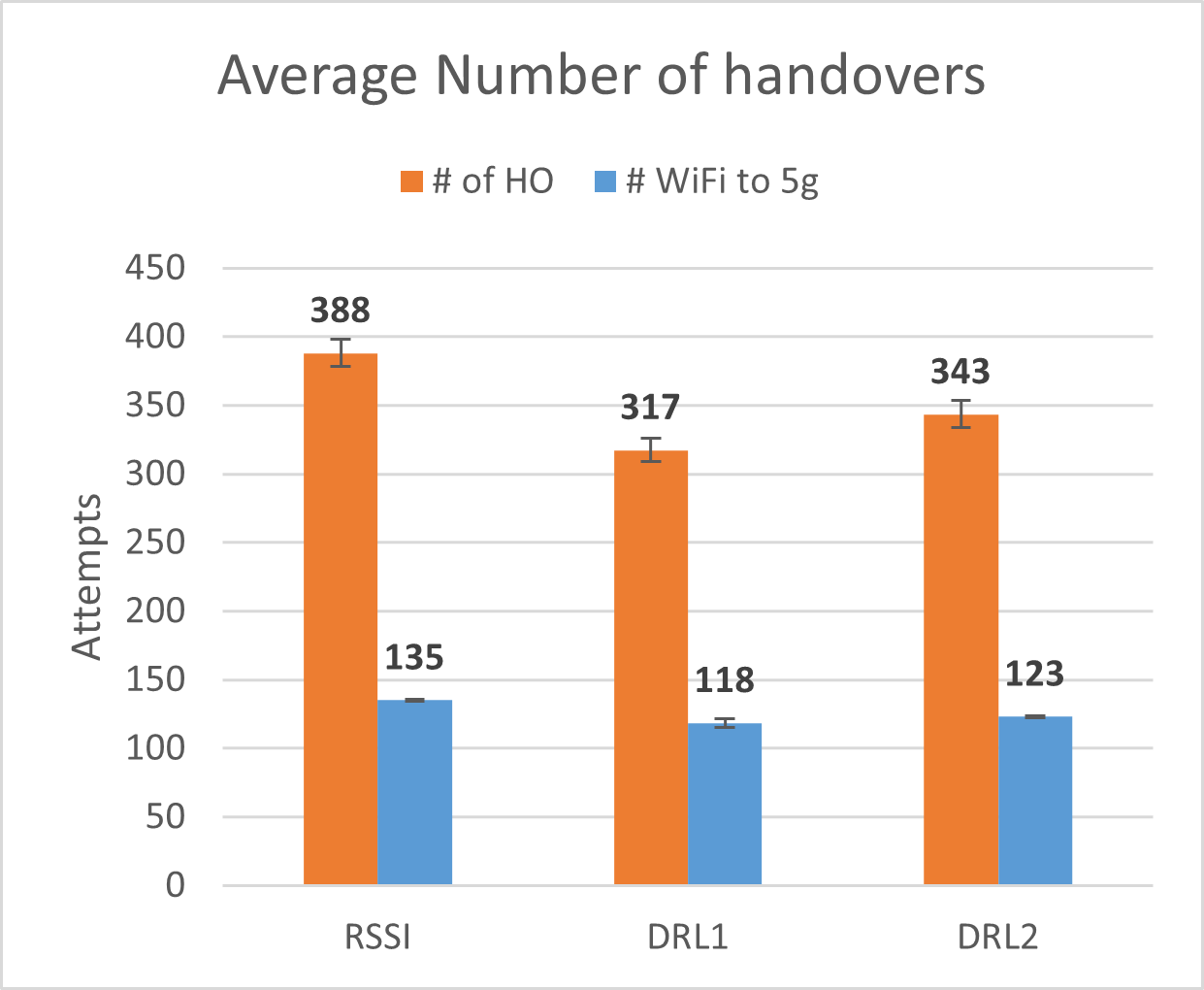}}
\caption{Number of Handovers}
\label{num_handovers}
\end{figure}

\subsection{Number of Handovers}
\label{subsect:numberofhandovers}
Fig.\ref{num_handovers} provides insights into the total number of handovers experienced by UEs as well as the number of handovers that moved UEs from WiFi to the gNB-like emulated node. The baseline traditional RSSI approach reveals not only a larger number of handovers, but also a greater tendency to handover the UE toward the 5G side of the emulation, where the gNB-like node has higher transmit power and is disproportionately favored over smaller cells, such as WiFi. ReFlexMRS reduces the total number of handovers in comparison to the RSSI-based handover, by 18\% for DRL1 and by 12\% for DRL2. This is due to the optimization of RSSI and delays in the case of DRL1 and DRL2, as well as consideration of network specific thresholds and multi-RAT preferences, e.g., WiFi over 5G. The proposed DRL agents also reduce the number of handovers from WiFi into the 5G side of the emulation, DRL1 by 13\% and DRL2 by 6\%, while balancing RSSI, delay, and throughput, as discussed next. The reported handover counts are decision level outcomes in the emulated setup and do not directly measure standardized handover failure rate, interruption time, or ping-pong definitions from full protocol stack systems.

The standard deviations observed for the total number of handovers are 19.57, 17.15, 19.77 for the RSSI baseline, DRL1, DRL2 handovers respectively. In the case of WiFi to 5G handovers, the standard deviations observed are 1.4, 6.8, 1.3 for the RSSI baseline, DRL1, DRL2 handovers respectively. 
The standard deviation observed in both cases is proportionally low, indicating the performance improvement of ReFlexMRS is consistent across multiple runs of the simulation. 

\begin{figure}[t]
\centerline{\includegraphics[width=\columnwidth]{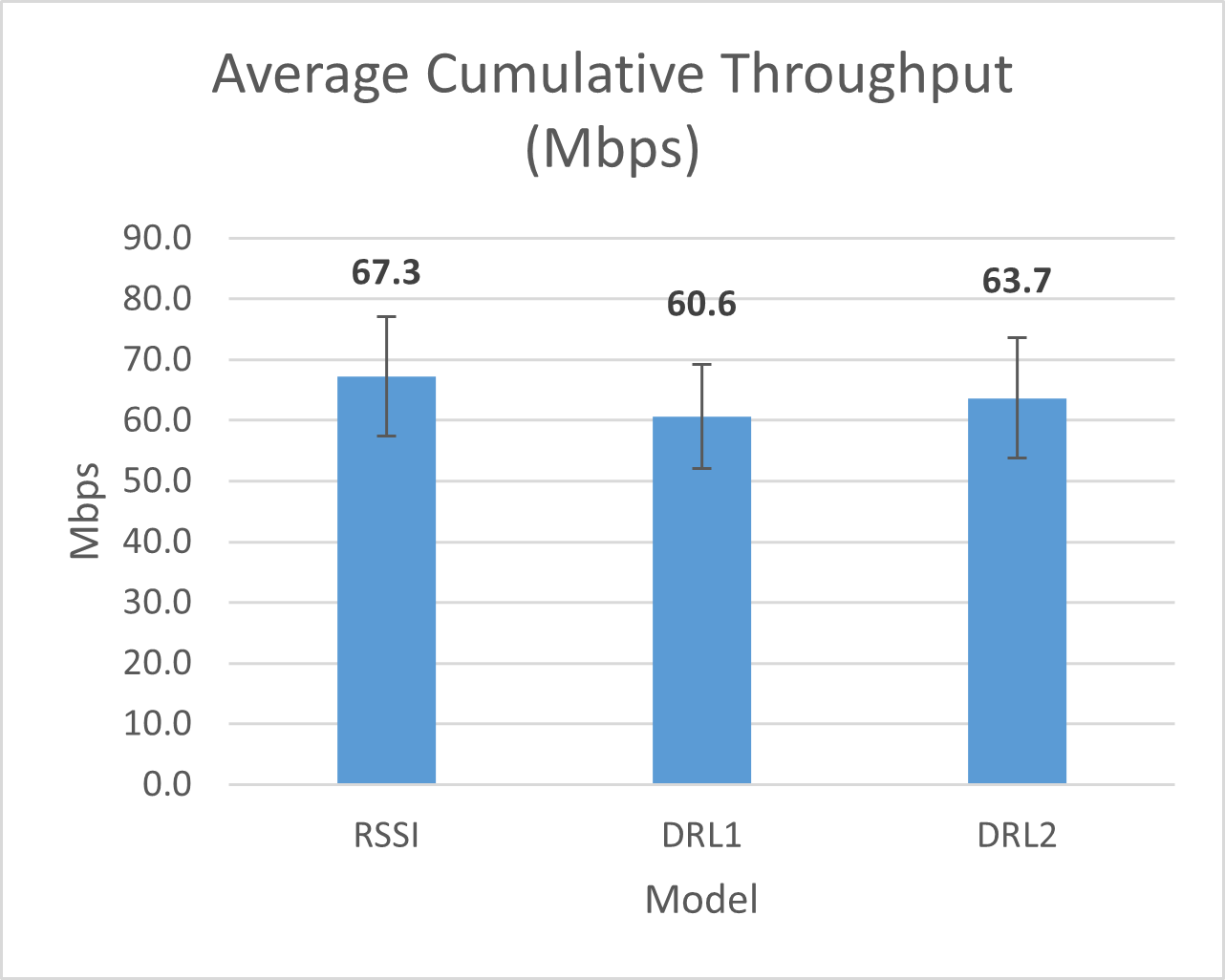}}
\caption{Average Throughput Comparison}
\label{throughput}
\end{figure}

\subsection{Throughput}
\label{subsec:throughput}
While the ReFlexMRS optimization techniques achieved improvement in various parameters, such as delay or selection of network types which is useful for congestion control or resource preservation, a high-priority criteria for future generations of networks is to maintain the throughput while balancing the other handover criteria. In this experiment, we collected the cumulative throughput obtained by the UE from the node it is connected to after each handover over the lifetime of the Mininet emulation, which is 60 minutes. We then calculate the average of these cumulative throughput values with respect to the total number of handover decisions made. This average is referred to as the average cumulative throughput.

Fig.\ref{throughput} shows the average cumulative throughput maintained by each technique, RSSI baseline, DRL1, and DRL2, for the duration of the Mininet WiFi emulation. The highest average cumulative throughput was exhibited by the RSSI baseline method, which is expected because the RSSI technique decision is  based on signal alone. While ReFlexMRS provides lower average cumulative throughput, it is purposely sacrificed to enable optimization of other possible handover criteria of interest. Our case study optimized delay and RAT-aware target selection while reducing unnecessary WiFi-to-5G transitions, thereby supporting WiFi offloading, while also considering throughput. Fig. \ref{throughput} shows that ReFlexMRS maintains approximately 90\% (DRL1) to 95\% (DRL2) of the RSSI approach's throughput. In addition, the standard deviation observed for the average cumulative throughput across each method is relatively low, indicating consistent performance across multiple runs of the simulation.
Thus, Fig.\ref{throughput} demonstrates that the new approach can achieve optimizations in other handover criteria while maintaining a comparable level of throughput.   

These results indicate that the proposed DQN-based handover models retain most of the RSSI-based baseline throughput while operating at substantially lower average delay per UE in the studied scenario. By incorporating delay and RAT selection preferences in addition to RSSI, ReFlexMRS expands the handover decision criteria and provides greater flexibility in selecting target networks under different QoS and offloading objectives.


\subsection{Key Takeaways}

The key takeaways comparing the baseline and DRL-based methods are as follows:

\begin{itemize}
    \item Throughput Trade-off: The RSSI-based approach achieves the highest throughput among the evaluated methods, but it does so at the expense of latency. ReFlexMRS retains approximately 90-95\% of the baseline throughput while achieving substantially lower link delay.
    \item Handover Stability: DRL1 and DRL2 reduce the total number of handovers by 18\% and 12\%, respectively, suggesting more stable associations and lower handover overhead.
    \item WiFi-to-5G Handover Reduction: DRL1 and DRL2 reduce WiFi-to-5G handovers by 13\% and 6\%, respectively, avoiding unnecessary transitions toward the higher power 5G node and preserving opportunities for WiFi offloading.

    \item  DRL1-DRL2 Trade-off: DRL1 achieves larger mean delay and handover reductions, whereas DRL2 retains more throughput and provides more consistent link delay outcomes, illustrating the different trade-offs produced by the combined feature and threshold-based formulations.

\end{itemize}

The following discussion interprets how the state and reward designs may have shaped these trade-offs and then positions ReFlexMRS relative to other DRL-based handover approaches.

\section{Discussion}
\label{sec:discussion}

\subsection{Impact of State and Reward Formulations}

\begingroup
Considering the three metric categories jointly, neither DRL formulation is uniformly superior. DRL1 produces stronger mean improvements in link delay and larger reductions in total and WiFi-to-5G handovers, whereas DRL2 retains more throughput and exhibits substantially lower variability in link delay difference. The two formulations therefore represent different operating trade-offs within ReFlexMRS rather than a single formulation dominating across all evaluated objectives.

We infer that these differences arise from the state and reward formulations used by the two agents. DRL1 combines RSSI and inverted delay into a single multiplicative QoS feature and applies a binary reward when the selected target maximizes that joint value. The multiplicative coupling favors candidates with a strong combined RSSI and delay value and can reduce the influence of a candidate that is advantageous primarily in RSSI, such as the higher power gNB-like node. However, DRL1 does not impose separate improvement margins for RSSI and delay or an explicit cost for performing a handover. Its learned operating point therefore depends on the relative scaling and distribution of both measurements in the collected observations. This behavior is consistent with the stronger average delay improvement, lower retained throughput, and greater link delay variability observed for DRL1. Its lower mean handover count should be interpreted as an emergent outcome of the learned policy, rather than as the direct effect of an explicit handover penalty.

DRL2 instead represents RSSI and inverted delay separately and applies explicit improvement thresholds and RAT-aware reward conditions. This structure allows the reward to distinguish changes in the two measurements and to encode the deployment preference for WiFi offloading when the configured QoS conditions are satisfied. The resulting reward shaping is consistent with DRL2's greater throughput retention and lower link delay variability, although its mean delay and handover reductions are smaller than those of DRL1. Because the two agents differ in both their state representations and reward functions, the present experiments cannot isolate the causal contribution of each design component. A component level ablation study would be required to determine the individual effects of feature combination, thresholding, and RAT-aware reward shaping.
\par\endgroup

\subsection{Comparison with Other DRL-Based Handovers}

Previous studies have applied RL to multi-RAT network selection, resource allocation, and traffic steering, while other studies have applied DRL to handover optimization within individual cellular or WLAN environments. ReFlexMRS differs by focusing on mobility-driven target selection between 5G and WiFi access nodes and by jointly evaluating signal quality, link delay, throughput, total handovers, and the direction of cross-RAT handovers. Because the related studies use different RATs, mobility models, optimization objectives, and performance baselines, a direct numerical comparison is not appropriate. We therefore provide a qualitative comparison of their objectives and reported performance.

Yacheur \textit{et al.}~\cite{yacheur2023drlrat} use DRL-based communication mode selection in a hybrid vehicular network using ITS-G5 and C-V2X, with the objective of improving reliability while limiting resource consumption. Although their method considers heterogeneous access technologies and application requirements, it does not evaluate 5G-WiFi mobility-driven handover frequency or offloading from cellular networks to WLANs. Allahham \textit{et al.}~\cite{allahham2022multiagent} jointly address dynamic network selection and RAN resource allocation in heterogeneous multi-RAT networks using deep multi-agent RL. Their work considers multiple QoS, energy, latency, cost, and resource allocation objectives, but network association and resource allocation are the primary objectives rather than mobility-driven handover stability.

De Santis \textit{et al.}~\cite{desantis2022satellite} apply DQN-based network selection across 4G, 5G, and satellite access points. Their evaluation emphasizes connection acceptance, load balancing, and resource allocation. Habib \textit{et al.}~\cite{Habib2023trafficsteering,habib2023hierarchical} use DRL-based traffic steering between LTE and 5G NR, considering load and different traffic QoS requirements. In contrast, ReFlexMRS considers association changes caused by user mobility between 5G and WiFi nodes and explicitly evaluates link delay, total handovers, and the direction of WiFi-to-5G handovers.

Authors of \cite{yajnanarayana2020a} propose an RL-based handover in a standalone 5G mmWave network, where handover is modeled as a contextual multi-armed bandit problem solved using Q-learning. The proposed model is shown to maximize the average link beam gain, which is indirectly correlated to increased throughput, but did not provide comparable throughput results for comparison.

Voigt \textit{et al.}~\cite{gu2024drl5g} propose a policy-gradient-based PPO agent for handover optimization in standalone 5G mmWave networks. Their formulation of handover as an MDP is similar to our approach. The PPO agent achieved a data rate improvement of 2\% to 7.5\%, maintained similar handover failures, and increased ping-pong effects compared with a 3GPP baseline. ReFlexMRS instead accepts a small throughput reduction to improve delay, reduce handovers, and encourage selection of WiFi access points when the configured QoS conditions are satisfied.

Han \textit{et al.}~\cite{han2019a} propose a Deep Q-Convolutional Recurrent Network (DQCRN) for handover in a standalone SDN-based WLAN. Their MDP uses SINR levels from each AP as states, AP selection as the action, and uplink throughput as the reward. Their simulation uses eight WiFi APs and a Random Waypoint mobility model, whereas our case study uses both a gNB-like emulated node and WiFi APs with the Random Direction mobility model. The UE speeds are comparable: Han \textit{et al.} use 0.4 meters per second, while our UEs move at 1.44 kilometers per hour (approximately 0.4 meters per second). Han \textit{et al.} report a 3.8\% average throughput improvement over an RSS-based scheme, but do not consider cross-RAT selection, link delay, or handover direction. ReFlexMRS instead trades a small amount of throughput for lower delay, fewer handovers, and WiFi offloading.

\subsection{Offline Training, Emulation Scope, and Threats to Validity}

Mininet-WiFi was selected because it is an open-source emulator that supports SDN-controlled WiFi experimentation and can be customized to represent a 5G gNB-like node. However, the environment does not natively implement a 3GPP-compliant 5G NR MAC/PHY or the complete standardized handover protocol stack. The evaluation therefore targets the handover decision layer in a controlled multi-RAT emulation. It does not directly reproduce all standardized signaling exchanges, interruption timing, handover failure procedures, or native lower-layer 5G behavior. The reported results should consequently be interpreted as evidence of ReFlexMRS framework feasibility and QoS trade-off behavior in the evaluated scenario rather than deployment-scale validation.

The approach uses offline training based on previously collected emulation data because Mininet-WiFi does not provide the low-latency native integration required for online interaction with the DRL agent. This choice simplifies integration but prevents the agent from observing closed-loop outcomes such as handover success or failure during training. Likewise, the reported handover counts represent target selection decisions in the emulated setup and are not direct measurements of standardized handover failure rate (HFR), interruption time, or ping-pong handovers. Future environments with native multi-RAT protocol stacks and real-time agent interaction would enable these additional metrics and support broader validation.

\section{Conclusion and Future work}\label{sec_conclusion}

In this work, we presented ReFlexMRS, a configurable DRL framework for multi-RAT handover target selection using multiple network criteria. The framework jointly considers received signal strength and average link delay and provides both network independent and deployment specific state and reward formulations. The deployment specific formulation demonstrates how RAT characteristics and operator preferences can be incorporated to encourage traffic offloading toward WiFi access points when the configured signal quality and delay conditions are satisfied. Together, the two formulations demonstrate how the same DRL framework can be adapted according to the network information and operator knowledge available in a deployment, including the use of RAT-specific preferences to support WiFi offloading and reduce unnecessary reliance on higher power cellular nodes.

ReFlexMRS was evaluated against conventional RSSI-based selection in an emulated 5G-WiFi environment. The results show that the proposed formulations substantially reduce link delay, total handovers, and WiFi-to-5G handovers while retaining approximately 90--95\% of the baseline throughput. Thus, ReFlexMRS improves the balance among competing QoS and mobility management objectives rather than maximizing throughput alone. Future work will investigate online training and native multi-RAT simulation or emulation environments that support real-time interaction between the DRL agent and 5G, WiFi, and other RAT protocol stacks.

\bibliographystyle{unsrt}
\bibliography{references.bib}

\end{document}